\documentstyle[11pt]{article}

\begin{document}
\thispagestyle{empty}

\begin{center}
               RUSSIAN GRAVITATIONAL SOCIETY\\
               INSTITUTE OF METROLOGICAL SERVICE \\
               CENTER OF GRAVITATION AND FUNDAMENTAL METROLOGY\\

\end{center}
\vskip 4ex
\begin{flushright}
                                         RGS-CSVR-006/95
                                         \\ gr-qc/9510039

\end{flushright}
\vskip 15mm

\begin{center}
{\large\bf Comments on ''The Principle of Self-Consistency as a consequence \\
 of the  Principle of Minimal Action'' } \\

\vskip 5mm
{\bf M.Yu. Konstantinov }\\
\vskip 5mm
     {\em VNIIMS, 3-1 M. Uljanovoj str., Moscow, 117313, Russia}\\
     e-mail: konst@cvsi.rc.ac.ru \\
\end{center}
\vskip 10mm

\begin{abstract}
The so called ''Principle of the self-consistency'' for space-time models
with causality violation, which was firstly formulated by I.D.Novikov, is
discussed for the test particle motion and for test scalar field. It is
shown that the constraints, which provide the self-concistensy of test
particle motion have pure geometrical (topological) nature. So, the recent
statement that ''The Principle of self-consistensy is a consiquence of the
Principle of minimal action'' is wrong.
\end{abstract}

\vskip 10mm

\vskip 30mm

\centerline{Moscow 1995}
\pagebreak


\section{Introduction}

In the recent years the big attention were paid to the space-time models
with closed time-like curves (causality violation)~\cite{mthyu}-\cite{krama}%
. As a rule the most authors try to find physical laws which may forbid the
creation of closed time-like curve and the existence of such models~\cite
{kim}-\cite{tanaka}. The declarations that in the vicinity of closed
time-like curves standard methods of differential geometry and general
relativity do not work~\cite{deutsch} as well as statements about
inconsistency of physics in the presence of time machine~\cite{krama} were
made also. The alternative declarations were made also. In particular, as it
was pointed out in~\cite{idn}, events on closed time-like curves should
causally influence each other along the loops in time in a self-adjusted,
consistent way. This requirement was originally formulated as the
''Principle of self-consistency'' according to which the Cauchy data for
space-time models with causality violation must be completed by additional
self-consistency conditions~\cite{idn}. The particular form of such
conditions for test particle in the simple non-causal space-time model was
analyzed recently in~\cite{novikov95}.

The impossibility of time machine creation as a result of some dynamical
process was shown recently in~\cite{myuk92}-\cite{myuk95}. In particular, it
was shown that in according with theorems about global hyperbolicity and
Cauchy problem~\cite{hawkell}-\cite{choquet} the models with causality
violation could not be considered as the result of dynamical evolution of
some initial space-like configuration and must be considered as a solution
of some boundary problem. It was pointed out also that the so called
''principles of self-consistency'' \cite{idn} follows directly from the
definition of fields on space-time. Nevertheless in paper \cite{novikov95}
it was declared that the so called ''principles of self-consistency'' is the
direct consequence of the principle of minimal action. By this reason in
this note we analyse in details the same model of the test particle in
non-causal space-time which was considered in \cite{novikov95}. The next
section contains the brief description of the model, and in the section 3
the role of the constraints and the action principle (equations of motion)
is analyzed. It will be shown that the conditions, which were initially
called in \cite{idn} as the ''principles of self-consistency'', are really
the constraints which follow from the geometry of the model. This result
confirm both our previous notions \cite{myuk92}-\cite{myuk95} and the
statement of \cite{novikov95} that the ''principles of self-consistency'' is
not independent assumption. It shows also that the statement of \cite
{novikov95} that ''the ''principles of self-consistency'' is the direct
consequence of the principle of minimal action'' is wrong. The constraints
equations for test scalar field are briefly considered in section 4. The
last section contains some concluding remarks.

\section{The model for the test particle}

Following to \cite{novikov95} we consider the non-relativistic motion of
self-interacting test particle of mass $m$ in the background with a wormhole
''time machine''. The exterior space-time is supposed to be Minkowskian and
the sizes of the wormhole mouths are negligibly small (point-like mouths'
approximation) and at rest in some reference frame. For definiteness, we
shell assume that the mouths of the wormhole in the exterior space-time have
coordinates $(t,\vec r_A)$ and $(t+\tau ,\vec r_B)$, where $\tau =\tau (t)>0$
and $\tau (t)\geq \vec r_B-\vec r_A$ for $t_1\leq t\leq t_2$. So, the region
$t_1\leq t\leq t_2+\tau (t_2)$ of the exterior space-time contains the paths
of closed time-like or null curves which violate causality (we use geometric
units where $c=1$).

As in \cite{novikov95}, the following particle's motion will be considered.
The particle starts at time $t_i$ in the position $\vec r_i$, enters to the
mouth (B) of the wormhole at time $\overline{t}+\tau (\overline{t})$
(position $\vec r_B$), where $\overline{t}>t_1$, exits from the other mouth
(A) at the earlier time $\overline{t}$ (position $\vec r_A$) and finally
ends its trajectory at time $t_f$ in the position $\vec r_f$. The path
length of the wormhole handle is assumed to be infinitely short, so the
motion through the wormhole in the proper time of the particle happens
almost simultaneously. According to an external observer, instead, the
particle traversing the time machine travels back in time by the amount $%
\Delta t=-\tau (\overline{t})$ where by definition $\overline{\tau }(%
\overline{t})>0$ and $\overline{\tau }(\overline{t})\geq \vec r_B-\vec r_A$
by assumption\footnote[1]{%
The last condition ($\tau \geq \vec r_B-\vec r_A)$ was not mentiond in\cite
{novikov95}, but it is the necessary condition for the existence of the
closed time-like curves.}. For simplicity, the motion with the only
self-intersection of the particle world line in the point with coordinates $%
(t_0,\vec r_0)$ will be considered in the following.

\section{Self-consistency conditions and the principle of minimal action}

Consider the region of exterior Minkowskian space-time with $\overline{t}<t<%
\overline{t}+\tau (\overline{t})$. The world line of the particle may be
considered in this region as two world lines of two copies of the same
particle with positions $\vec r_1(t)$ and $\vec r_2(t)$. Both particles may
be considered as independent objects which interact by means of potential $V$
of special type. The motion of such system is described by the action (term $%
S_{12}$ in equation (4) of \cite{novikov95})
\begin{equation}
\label{action}S=\int\limits_{\overline{t}}^{\overline{t}+\tau }dt\left\{
\frac m2\dot{\vec r}_1^2(t)+\frac m2\dot{\vec r}_2^2(t) - V(\left| \vec
r_1(t)-\vec r_2(t)\right| )\right\}
\end{equation}
The standard variation of $S$ with respect to $\vec r_1(t)$ and $\vec r_2(t)$
gives the motion equations for both particles (eqns. (6) of \cite{novikov95}%
).

The geometry of the model imposes some additional limitations on the
possible motion. Namely, the entrance of the particle into mouth B and its
exit from A may be written formally as (eqns. (3) of \cite{novikov95})
\begin{equation}
\label{ent}\vec r_1(\overline{t}+\tau (\overline{t}))=\vec r_B
\end{equation}

\begin{equation}
\label{exit}\vec r_2(\overline{t})=\vec r_A
\end{equation}
and the self-intersection of the particle world line has the form (eqns.
(14) of \cite{novikov95})
\begin{equation}
\label{intersect}\vec r_1(t_0)=\vec r_2(t_0)=\vec r_0
\end{equation}

In \cite{novikov95} an exact solution of the motion equations, which
correspond to the action (\ref{action}), with the constraints (\ref{ent})-(%
\ref{intersect}) was obtained. It was stated that the existence of such
solution, which minimize the action functional, shows that the ''Principle
of self-consistency'' is a consequence of the ''Principle of minimal
action'' \cite{novikov95}.

Let's analyse this statement in more details. To this end consider the
constraint equations (\ref{ent})-(\ref{intersect}). The condition (\ref{ent}%
) states that particle fall down into wormhole at some time $\overline{t}%
+\tau (\overline{t})$ independently from the previous history and the
condition (\ref{exit}) states that the world line $(\overline{t},\vec r_2(%
\overline{t}))$ is the continuation of the world line of the same particle.
Therefore according to the constraints (\ref{ent})-(\ref{intersect}) the
points ($t,\vec r_1(t))$ and $(t,\vec r_2(t))$ are the points of different
paths of the world line of the same particle. By the same reason the point $%
(t_0,\vec r_1(t_0))=(t_0,\vec r_2(t_0))=(t_0,\vec r_0)$ is the
self-intersection point of the same particle world line. Moreover,
conditions (\ref{ent})-(\ref{intersect}) state also, that the
self-intersection of the particle's world line does not prevent its passing
through the time machine. Hence, the conditions (\ref{ent})-(\ref{intersect}%
) prevent the appearance of the so-called ''Polchinski paradox'' \cite{krama}
or the ''grandmother paradox'' which are usually associated with the
existence of the time machine.

So, the self-consistency of the solution (the absence of some ''paradoxes'')
is provided not by the ''Principle of minimal action'', but by the
constraint equations (\ref{ent})-(\ref{intersect}). The geometrical (to be
more precise, topological) nature of these equations is obvious. By this
reason, namely the constraints (\ref{ent})-(\ref{intersect}) may be
naturally called as the ''self-consistency'' conditions.

Of cause, the constraints (\ref{ent})-(\ref{intersect}) are closely
connected with the motion equations. Namely, the number of the particle's
entrance into the time machine, the existence and number of
self-intersections of its world line, as well as the exact values of the
parameters $\overline{t}$, $t_0$ and $\vec r_0$ for every self-intersection
are the subject of the motion equation. But if the world line of the
particle has self-intersection, the local solution of the motion equations
near each point of self-intersection must satisfy to the constraints (\ref
{ent})-(\ref{intersect}) or their generalization.

\section{Test scalar field in non-causal two dimensional space-time}

For the sake of simplicity, consider the test scalar field $\varphi $ in the
following two-dimensional space-time model. As above, the exterior
space-time will be assumed to be flat Minkowskian space with coordinates $%
(t,x)$ and the wormhole connects the points with coordinates $(t,0)$ and $%
(t+\tau ,l)$, where $\tau $, $l$ are some constants. The length of the
wormhole is supposed to be infinitesimally small. So, the condition of
causality violation is $\tau >l$.

In the exterior space-time of such model scalar field $\varphi $ satisfies
to the usual wave equation
\begin{equation}
\label{wave}\varphi _{tt}-\varphi _{xx}=0
\end{equation}
with constraint
\begin{equation}
\label{constr}\varphi (t,0)=\varphi (t+\tau ,l)=u(t)
\end{equation}
It is easy to see that these constraint in general cannot be reduced to the
standard boundary conditions for wave equation in spatially bounded region
even in the models without causality violation.

The exact solutions of the equations (\ref{wave}) with constraint (\ref
{constr}) may be easily found using the standard separation of variables in
two particular cases:

{\bf (1)} $u(t)=0$. In this case the interior and exterior solutions are
independent as well as the solutions in the regions $x\leq 0$, $0\leq x\leq
l $, and $x\geq l$.

{\bf (2)} $\tau =2\pi n$, where $n>l/(2\pi )$ in the causality violation
case. In this case solution of the problem (\ref{wave})-(\ref{constr})
coincides with the well known solution of (\ref{wave}) with boundaries
conditions
$$
\varphi (t,0)=\varphi (t,l)=u(t)
$$

Consideration of the problem (\ref{wave})-(\ref{constr}) in the general case
as well as the consideration of the finite length of the wormhole is the
subject of separate paper. Nevertheless, it may be shown that nontrivial
solutions exist only if parameters $\tau $ and $l$ satisfy to some
constraint.

\section{Discussion}

So, we have considered the problem of self-consistency for the motion of
classical test particle and classical test scalar field in the space-time
with Lorentzian wormhole, which violate causality. It was shown, that in the
case of the test particle the self-consistency of the solution (the absence
of some ''paradoxes'') is provided not by the ''Principle of minimal
action'' as it was stated in \cite{novikov95}, but by the constraint
equations (\ref{ent})-(\ref{intersect}). By this reason, namely the
constraints (\ref{ent})-(\ref{intersect}) may be naturally called as the
''self-consistency'' conditions in the test particle case. For test scalar
field the self-consistency of the solution is provided by the constraints (%
\ref{constr}). It is obvious that these constraints have geometrical (to be
more exact, topological) nature: in the case of test particle the
constraints (\ref{ent})-(\ref{intersect}) follow from the definition of the
line on manifold and the constraints (\ref{constr}) are the part of the
definition of function on manifold.

Of cause, the above result does not mean that the ''self-consistency''
conditions have no connection with the action principle. The constraints (%
\ref{ent})-(\ref{intersect}) and (\ref{constr}) contain parameters whose
values depend from the motion and field equations. Namely, the particle's
entrance into the time machine and the existence of self-intersection of its
world line, as well as the exact values of the parameters $\overline{t}$, $%
t_0$ and $\overrightarrow{r}_0$ are the subject of the motion equation. But
if the world line of the particle has self-intersection, the local solution
of the motion equations near the point of self-intersection must satisfy to
the constraints (\ref{ent})-(\ref{intersect}) or their generalization.
Similarly, for the test scalar field parameters $\tau $ and $l$ (more
generally, $\tau $, $x_A(t)$ and $x_B(t+\tau )$) are connected by the
equation (\ref{wave}). Additional restriction on the parameters appears in
the quantum case, whose detailed consideration as well as the consideration
of more general models, is the subject of separate paper.

\end{document}